\documentclass[preprint,12pt,3p,times,authoryear,nopreprintline]{elsarticle}

\usepackage{lineno}
\usepackage{hyperref}
\usepackage{soul}
\usepackage{tabularx}
\usepackage{array}
\usepackage{epsfig}
\usepackage{ulem}
\usepackage{xcolor}
\usepackage{tikz}
\usepackage{pgf-pie}
\usepackage{caption}
\usetikzlibrary{shapes.geometric,patterns}
\usepackage{pgfplots}
\pgfplotsset{compat=1.17} 
\usepackage{booktabs}
\usepackage{multirow}
\usepackage{amssymb}
\usepackage{amsmath}
\usepackage{amsthm}
\usepackage{graphicx}
\usepackage{rotating} 
\usepackage{caption}
\usepackage{booktabs} 
\usepackage{longtable} 
\usepackage{multirow}
\usepgfplotslibrary{colorbrewer}
\usepackage{booktabs} 
\usepackage{caption} 
\usepackage{graphicx} 
\usepackage{hyperref} 
\usepackage{threeparttable} 
\usepackage{tablefootnote}
\usepackage{lipsum} 
\usepackage{float} 
\usepackage{longtable} 
\usepackage{afterpage}

\journal{Computers \& Education}

\begin{document}

\begin{frontmatter}

\title{The Status Quo and Future of AI-TPACK for Mathematics Teacher Education Students: A Case Study in Chinese Universities}


\author[GXNU]{Meijuan Xie}

\author[GXNU]{Liling Luo\corref{corref1}}
\ead{593929431@qq.com}

\cortext[corref1]{Corresponding author.}
\affiliation[GXNU]{organization={School of Mathematics and Statistics, Guangxi Normal University},
            city={Guilin},
            postcode={541004}, 
            country={China}}

\begin{abstract}
As artificial intelligence (AI) technology becomes increasingly prevalent in the filed of education, there is a growing need for mathematics teacher education students (MTES) to demonstrate proficiency in the integration of AI with the technological pedagogical content knowledge (AI-TPACK). To study the issue, we firstly devised an systematic AI-TPACK scale and test on 412 MTES from seven universities. Through descriptive statistical analyses, we found that the current status of AI-TPACK for MTES in China is at a basic, preliminary stage. Secondly, we compared MTES between three different grades on the six variables and found that there is no discernible difference, which suggested that graduate studies were observed to have no promotion in the development of AI-TPACK competencies. Thirdly, we proposed a new AI-TPACK structural equation model (AI-TPACK-SEM) to explore the impact of self-efficacy and teaching beliefs on AI-TPACK. Our findings indicate a positive correlation between self-efficacy and AI-TPACK. We also come to a conclusion that may be contrary to common perception, excessive teaching beliefs may impede the advancement of AI-TPACK. Overall, this paper revealed the current status of AI-TPACK for MTES in China for the first time, designed a dedicated SEM to study the effect of specific factors on AI-TPACK, and proposed some suggestions on future developments.


\end{abstract}


\end{frontmatter}

\section{Introduction}
\label{sec1}
The technological pedagogical content knowledge (TPACK) has undergone further development and now is a significant component of the criteria used to evaluate the competence of educators \citep{ARCHAMBAULT2010TPACK}. The rapid development of Artificial Intelligence (AI) has had a significant impact on various aspects of education \citep{XU2021ai}. In response to this, TPACK framework has been adapted to incorporate AI elements, leading to the emergence of the Artificial Intelligence Integrated Technology Pedagogy Content Knowledge (AI-TPACK) framework. This consequently imposes specific obligations on educators with regard to the utilisation of AI technology \citep{park2023integrating}. The advent of this concept represents a significant advancement in the domain of educational technology, offering educators a novel lens through which to assess and enhance their pedagogical approaches.  The emphasis on AI-TPACK in teacher training programs is leading to the advancement of AI-related competencies, including AI-Technological Knowledge (AI-TK), AI-Technological Content Knowledge (AI-TCK), and AI-Technological Pedagogical Knowledge (AI-TPK), collectively enhancing the overall level of AI-TPACK. The emergence of AI-TPACK not only presents a new path for educators to advance their professional development, but also provides students with more expansive and adaptable manners for learning. It is a powerful force in the digitalisation process of promoting the modernisation of education \citep{CELIK2023teacher}.

The advent of AI-TPACK has introduced a novel set of challenges and requirements for educators. AI-TPACK emphasises  the necessity of contemplating not only the incorporation of TK, PK, and CK into the professional development of educators, but also the potential of AI to enhance a more personalised and efficacious learning experience. It necessitates that teachers possess proficiency in AI technology, demonstrate expertise in integrating it with conventional pedagogical approaches, and exhibit creativity in developing novel instructional strategies \citep{jiang2021using}.
\cite{Geng2021student} highlights the necessity for educators to maintain the advancement of their subject knowledge in order to facilitate a more comprehensive and multifaceted understanding for their students, with the assistance of AI. It is imperative that teachers adopt a lifelong learning approach and remain abreast of developments in AI in order to effectively navigate the evolving landscape of education \citep{an2023modeling}.

Teachers' self-efficacy can be defined as a subjective assessment of their own teaching ability \citep{bandura1997self}. It reflects their positive expectations of the teaching process and outcomes, which in turn influence their teaching behaviours and students' learning outcomes \citep{KLASSEN2014effect}. Teaching beliefs represent the fundamental tenets and values that educators internalize regarding the objectives of education, the learning process, the characteristics of learners, and the content and methodologies of teaching \citep{thurm2020effects}. Teaching beliefs constitute the foundation of a teacher's educational philosophy and practice, and they shape the teacher's educational conduct \citep{gilakjani2017teachers}. Teachers who exhibit high levels of teacher self-efficacy and teaching beliefs are more likely to demonstrate confidence in exploring and implementing elements of the AI-TPACK framework \citep{lee2010exploring}. Furthermore, they may demonstrate greater confidence in problem-solving when confronted with the challenges of integrating technology \citep{Hasan2023self,yildiz2023examining,lai2022beliefs,Chai2019beliefs}.

MTES play a pivotal role in the development of education, serving as the backbone of basic education and tasked with imparting mathematical knowledge, cultivating critical thinking skills, and stimulating student interest. They also have a significant responsibility to facilitate the modernization and digital transformation of education. It would be advantageous to integrate the teaching of AI technical knowledge from the school curriculum and development of AI-TK competence with PK and CK. This approach is more conducive to enhance the overall capability of AI-TPACK. The acquisition of AI-TPACK skills by MTES will result in a notable enhancement in their overall quality, an increase in their employability and a promotion of the sharing of educational resources. It is therefore important to prioritise and strengthen the cultivation of AI-TPACK skills among MTES in order to advance the development of education, achieve educational equity and foster the development of innovative talents. So in order to better enhance the AI-TPACK competence of MTES, the paper was based on three key research questions: 

\begin{itemize}
\item RQ1: What is the current status of AI-TPACK for MTES ?
\item RQ2: What are the differences in MTES between grades ?
\item RQ3: What are the effects of self-efficacy and teaching beliefs on AI-TPACK ?
\end{itemize}

\section{Literature Review}
\label{sec2}

\subsection{What is TPACK}
\begin{figure}[ht!]
\centering
\includegraphics[width=1.0\textwidth]{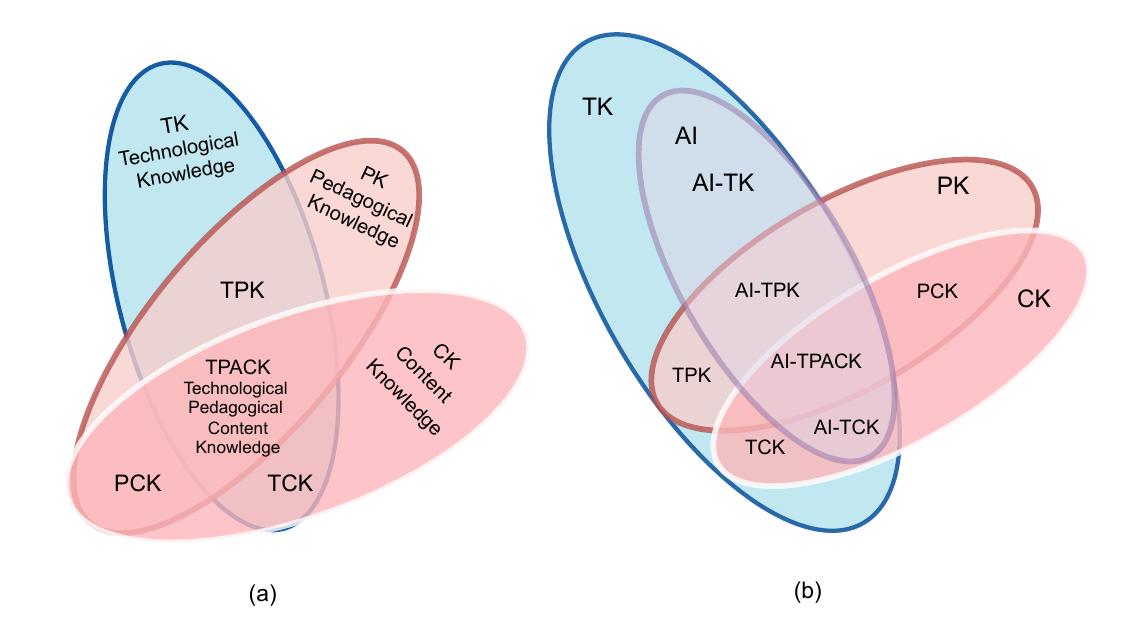}
\caption{TPACK and AI-TPACK framework.}
\label{TPACK}
\end{figure}

\cite{Shulman1986pck} proposed the Pedagogical Content and Knowledge (PCK) framework, which provided guidance to educators on the selection of appropriate teaching strategies and resources. This framework had a significant influence on the advancement of teacher education students and professional development \citep{Baumert2010mathpck}. Building on PCK framework and adapting to the educational requirements of the Information Age, \cite{Mishra2006TPACK} proposed the Technological Pedagogical Content Knowledge (TPACK) framework, which has since become a significant theoretical framework in the field of education \citep{SCHMID2024important}. As shown in Fig.\ref{TPACK} (a), TPACK describes a framework for teachers to combine Technological Knowledge (TK), Pedagogical Knowledge (PK), and Content Knowledge (CK) to facilitate effective teaching practices. Technological Pedagogical Knowledge (TPK) represents The combination of technology and pedagogy. Technological Content Knowledge (TCK) represents the interdisciplinary field concerned with the integration of technology and content. PCK represents the integration of pedagogical and content knowledge. The combination of these components constitutes the TPACK \citep{Mishra2006TPACK}.

Following the proposal of the TPACK framework, numerous scholars have engaged in the development and refinement of this conceptual model. As demonstrated in Table \ref{tab:TPACK Literature}, contextualisation is highlighted as a key consideration \citep{Kelly2008context}. Furthermore, it is closely associated with the field of teacher education \citep{Hofer2008teacher} and the devising of assessment instruments \citep{Groth2009assessment}.

\vspace*{-0.08cm}
\begin{table*}[!t]
 \caption{The development history of TPACK framework.}
 \label{tab:TPACK Literature}
    \centering
    \small
\scalebox{.95}
    {    
    \setlength{\extrarowheight}{5pt}
    \begin{tabular}{p{5.5cm}@{\hspace{1.5cm}} p{9cm}}
    \toprule
    \textbf{Literature} & \textbf{Contributions to the development}\\
    \midrule
    {{\cite{Kelly2008context, ANGELI2009context}}}& {They placed significant emphasis on the importance of context in the TPACK framework. They argued that teachers need to select suitable technological tools and pedagogical approaches for diverse contexts.} \\
    \midrule
    {{\cite{Ching2010tpack, TONDEUR2012teacher,Baran2019strategy, JIN2022teacher}}} & {They focused on TPACK framework be fully integrated into educational practices. And they also focused on teachers, which should include not only in-service teachers, but also those pre-service teachers and teacher education students.}\\
    \midrule
    {{\cite{Schmidt2009assessment, Groth2009assessment, Zelkowski2013assessment, AKYUZ2018assessment, YEH2021framework, aktacs2022assessing}}}  & {They developed tools for assessing teachers' TPACK capabilities. These assessments were conducted on all groups of teachers, and were furthermore applied to different disciplines, such as Math.} \\
    \bottomrule
    \end{tabular}
}
    \vspace*{0.1cm}
    \vspace*{-0.1cm}
\end{table*}

\subsection{AI in education} 
Artificial intelligence (AI) can be defined as the demonstration of intelligent behaviours by machines that are often similar to human behaviours, including learning, understanding language and creating \citep{minh2022explainable}. AI is revolutionising the world \citep{ZHANG2021wide}. The utilisation of AI in the educational sector is becoming increasingly prevalent, and it is transforming the conventional teaching paradigm and exerting a discernible influence on both educators and students \citep{Zhai2021review}. For students, \cite{Ali2024agency} identify a growing utilisation of AI-powered learning technologies. The evidence suggests that these technologies have a beneficial impact on student learning. These technologies are employed in a variety of ways. For example, AI has the capacity to automatically adjust the content and difficulty of learning according to the student's learning progress and style. It is also able to provide real-time feedback and guidance, and assist students in solving problems related to their learning \citep{HUANG2023pesonalized,Chen2021personalized}. For teachers, AI can facilitate access to more valuable information, enabling them to enhance their lessons \citep{Kim2020assistant}, obtain detailed feedback \citep{Hahn2021feedback}, and optimise their teaching efficacy \citep{TEO2021teaching}. The use of AI in education to tailor learning for each student can replace the one-size-fits-all approach. In addition, AI can provide a variety of assistive teaching tools, including smart drawing boards and chatbots, which could enhance student interest and facilitate more effective teaching practices for educators \citep{lo2024chat}. It is important to note that while AI offers numerous advantages, it also presents a number of potential challenges. The application of AI must adhere to a set of ethical principles, including those pertaining to fairness and transparency \citep{Budee2024ethical}. 

\subsection{Combination of AI and TPACK}
The Artificial Intelligence Integrated Technology Pedagogy Content Knowledge (AI-TPACK) framework was developed, which with the objective of adapting to the advancements in information technology \citep{luckin2022ai}. It incorporates elements of artificial intelligence on top of TPACK in order to facilitate the integration of these technologies in education (see Fig.\ref{TPACK} (b)). The integration of AI into TK, intertwining it with PK and CK to form AI-TPK, AI-TCK, and subsequently combining them with PCK to develop AI-TPACK represents a significant advancement in the field of knowledge management. In addition to utilising AI technology as a teaching aid, AI-TPACK also necessitates the profound integration of human instructors with such technology, thereby fostering a tripartite co-creation of technological, content knowledge, and pedagogical elements through a symbiotic blend of human and machine cognition \citep{CELIK2023teacher,cavalcanti2021automatic,SEUFERT2021teacher}. The Intelligent Tutoring System is employed by educators to provide personalised and precise tutoring to students, taking into account their level of knowledge and other personal circumstances \citep{Pai2021tutoring,MOHAMED2018tutoring}. Robot teachers serve as paraprofessionals, working in conjunction with educators to facilitate classroom instruction \citep{LUO2019CHATBOT}. They also provide feedback based on classroom records, which can assist human teachers in making more informed adjustments to their teaching methods \citep{Chocarro2023CHAT}. Furthermore, AI can be used in different fields, such as improving mathematical thinking skills \citep{frieder2023mathematical}, 
 assisting in language learning \citep{Jaeho2023english}. The advent of ChatGPT has enhanced the efficacy of educators in locating information and gaining a vast reservoir of knowledge in a relatively short time \citep{Emma2024app}.

The focus of AI-TPACK studies has been on in-service and pre-service teachers, with relatively little research conducted involving teacher education students. Nevertheless, teacher education students represent the backbone of future educators and play a pivotal role in educational advancement. It is worthy of note that the majority of current studies are conducted at a macro level, fail to subdivide subject knowledge, and lack relevance. In light of the aforementioned limitations of current research, this paper examines the current status of AI-TPACK competence among MTES, and the differences between universities and grades. Furthermore, two additional latent variables, namely self-efficacy and teaching beliefs, were incorporated into the analysis to ascertain their potential contribution to AI-TPACK.

\section{Methodology}
\label{sec3}
\begin{figure}[ht!]
\centering
\includegraphics[width=.8\textwidth]{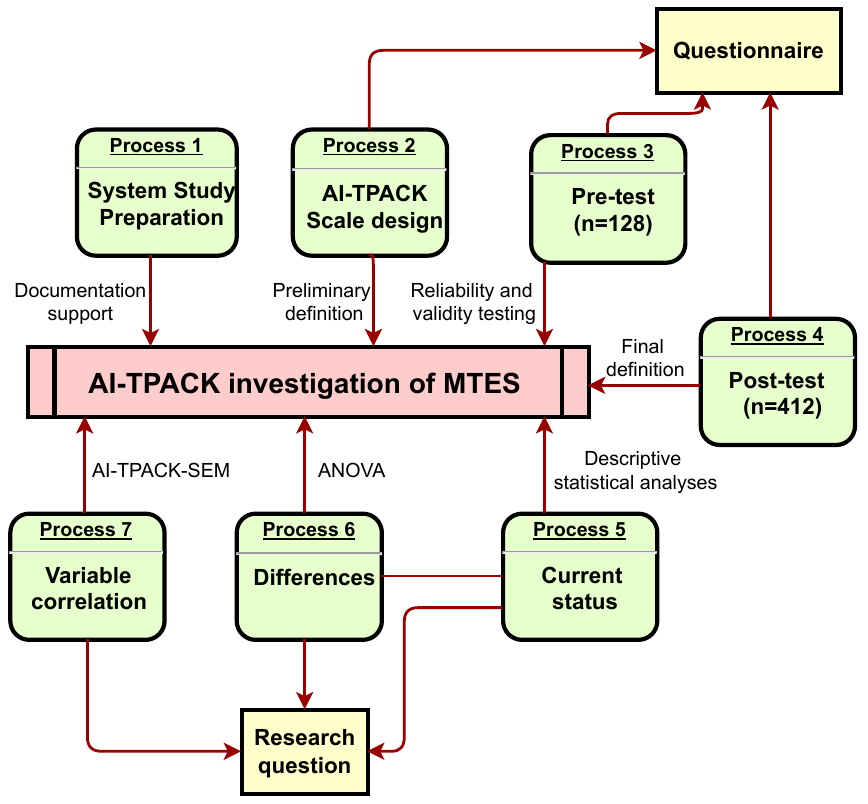}
\caption{MTES's AI-TPACK research framework.}
\label{framework}
\end{figure}

The overall framework of our paper are shown in Fig.\ref{framework}. This research is split into seven processes, and we will introduce they according to the following three parts: AI-TPACK scale design and refinement, participants, data mining and analysis. 

\subsection{AI-TPACK scale design and refinement}
\label{3.1}
 This section covers processes 1-4. At first, a systematic review and analysis of the relevant literature was conducted, and a preliminary AI-TPACK scale was designed. This questionnaire scale comprised of two parts, the former part investigates fundamental information regarding the students' backgrounds, including gender, educational institution, manners for acquiring knowledge about AI and the purpose of its applications.
And the later one delves deeper into the students' level on six variables in specialised areas, including AI-TK, AI-TCK, AI-TPK, AI-TPACK, self-efficacy, and teaching beliefs. In this scale, answer and scores of each variable of interest come from five items, and the options of each item is a 7-point Likert type ranged from “1: strongly disagree” to “7: strongly agree” \citep{likert1932technique}. Secondly, a pre-test is conducted on students of interest, and used Exploratory factor analysis (EFA) and Confirmatory factor analyses (CFA) to assess and modify the AI-TPACK scale. Our final revised AI-TPACK scale is shown in Table \ref{tab:scale}, it contained six variables and 24 items and each variable hold four items.

\subsection{Participants}
\label{3.2}
In the pre-test phase, our AI-TPACK scale was distributed online to senior MTES in one university. A total of 136 questionnaires were recovered, to eliminate disturbance factor and enhance data reliability, those completed in less than one minute were excluded, resulting in 128 valid questionnaires. These valid data come from 94 (73.4\%) females and 34 (26.6\%) males. During the Post-test phase, AI-TPACK scale was distributed online to MTES in seven similar universities. As with the preliminary survey, after excluding the invalid data, the final sample consisted of 412 valid questionnaires. The participants comprised 262 (63.6\%) senior university students, 105 (25.5\%) first-year graduate students and 45 (10.9\%) second-year graduate students. As to gender, 289 (70.1\%) of it were female and 123 (29.9\%) were male, this is also roughly the same distribution as in the pre-test data.

\subsection{Data mining and analysis}
\label{3.3}
\begin{figure}[ht!]
\centering
\includegraphics[width=0.58\textwidth]{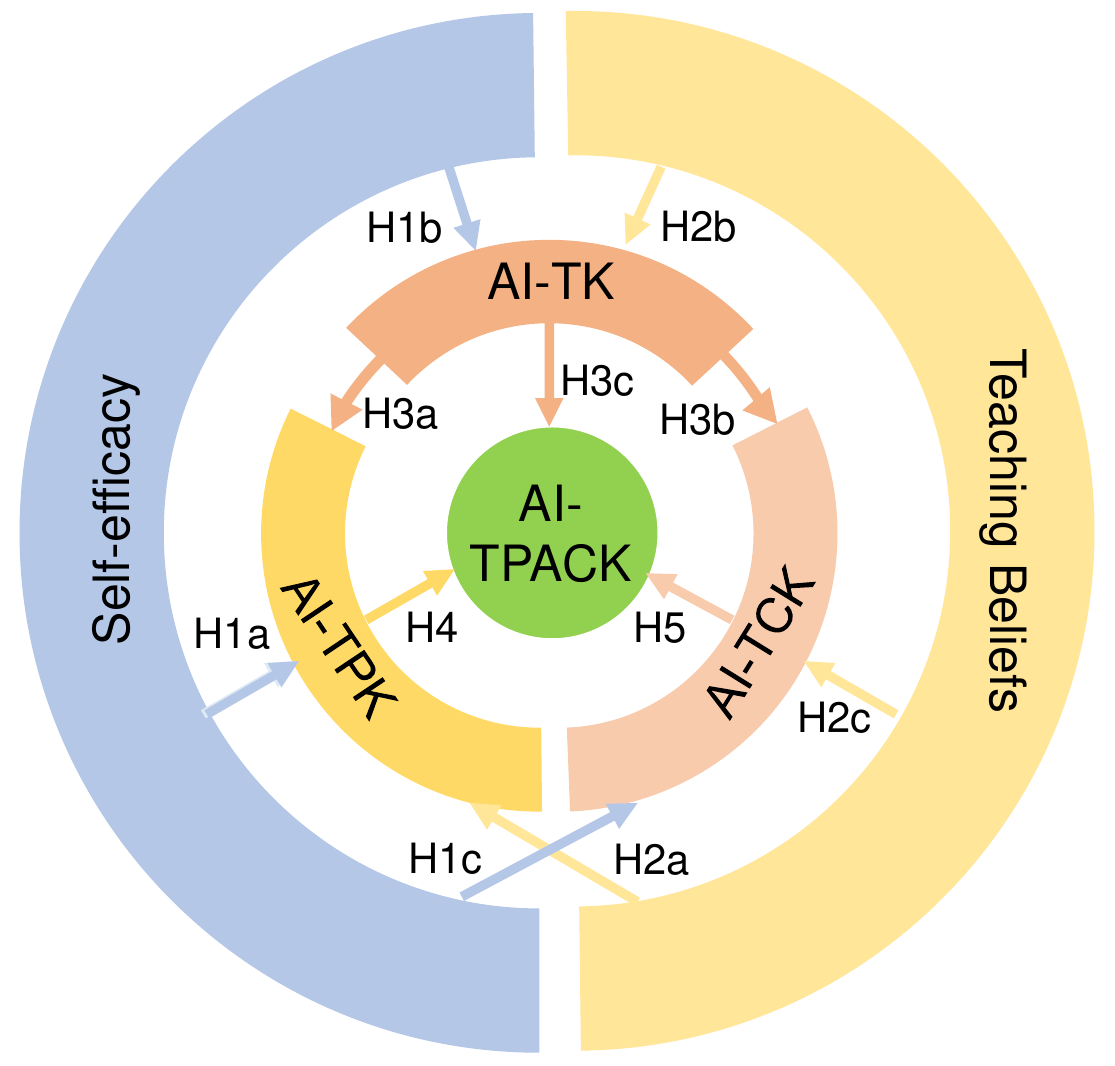}
\caption{Hypothetical paths of AI-TPACK-SEM.}
\label{hypothesize-circle}
\end{figure}

EFA and CFA were employed to assess the reliability and validity of the scale. Then we apply SPSS 26.0 for descriptive statistical to test the overall distribution of variables and a analysis of variance (ANOVA) to comparing differences in variables, in order to find out the current status of AI-TPACK for MTES, and the differences between grades. Additionally, our innovative introduced curve-fitting model to analyze the relationship between the six variables two-by-two. We also proposed a brand new AI-TPACK-SEM, as shown in Fig.\ref{hypothesize-circle}. In this model, we hypothesized the interactions between the variables, 11 hypotheses in total, detailed information and description please refer to Table \ref{tab: Hypotheses}. The proposed AI-TPACK-SEM is subjected to structural equation modeling (SEM) analysis and the coefficients were calculated to elucidate the interactions between AI-TPACK components and self-efficacy and teaching beliefs.

\vspace*{-0.08cm}
\begin{table*}[!ht]
 \caption{Interpretation based on hypothetical paths}
 \label{tab: Hypotheses}
    \centering
    \small
\scalebox{1.0}
    {    
    \setlength{\extrarowheight}{3pt}
    \begin{tabular}{p{2.5cm}@{\hspace{1.5cm}} p{8cm}}
    \toprule
    \textbf{Hypotheses}
     & Description
     \\
    \midrule
    \textbf{H1a} & Self-efficacy positively influences AI-TPK. \\
    \textbf{H1b} & Self-efficacy positively influences AI-TK. \\
    \textbf{H1c} & Self-efficacy positively influences AI-TCK. \\
    \textbf{H2a} & Teaching Beliefs positively influences AI-TPK. \\
    \textbf{H2b} & Teaching Beliefs positively influences AI-TK. \\
    \textbf{H2c} & Teaching Beliefs positively influences AI-TCK. \\
    \textbf{H3a} & AI-TK positively influences AI-TPK. \\
    \textbf{H3b} & AI-TK positively influences AI-TCK. \\
    \textbf{H3c} & AI-TK positively influences AI-TPACK. \\
    \textbf{H4}  & AI-TPK positively influences AI-TPACK. \\
    \textbf{H5}  & AI-TCK positively influences AI-TPACK. \\
    \bottomrule
    \end{tabular}
}
    \vspace*{0.1cm}
    \vspace*{-0.1cm}
\end{table*}

\section{Results}
\label{sec4}
\subsection{Pre-test}
\label{4.1}
A pre-test was administered to a sample of 128 senior MTES in one university, and the reliability of the initial AI-TPACK scale was subsequently analysed. EFA and CFA were employed to validate the AI-TPACK scale. 

\begin{figure}[b!]
\begin{tikzpicture}
    \begin{axis}[
    width = 15cm,
    height = 8.5cm,
        ylabel={Factor Loading},
        xlabel={Component},
        xmin=0, xmax=7,
        ymin=0.4, ymax=1,
        xtick={0,1,...,6},
        ytick={0.4,0.5,0.6,0.7,0.8,0.9},
axis line style={double=black, double distance=1pt}, 
 axis line style={->, black},
        axis lines*=left, 
    ]
    
    \addplot[
    color=red,
    only marks, 
    mark=* ,
     mark size=5pt
]  coordinates {
        (1,.797)
        (1, .755)
        (1, .557)
        (1, .466)
    };

     \addplot[
    color=black,
    only marks, 
    mark=* ,
     mark size=5pt
]  coordinates {
        (2, .771)
        (2, .578)
        (2, .656)
        (2, .688)
    };

     \addplot[
    color=purple,
    only marks, 
    mark=*  ,
     mark size=5pt
]  coordinates {
        (3, .598)
        (3, .593)
        (3, .550)
        (3, .661)
    };

     \addplot[
    color=cyan,
    only marks, 
    mark=*  ,
     mark size=5pt
]  coordinates {
        (4, .677)
        (4, .596)
        (4, .696)
        (4, .568)
    };

     \addplot[
    color=black,
    only marks, 
    mark=*  ,
     mark size=5pt
]  coordinates {
       (5, .514)
        (5, .673)
        (5, .806)
        (5, .679) 
    };

     \addplot[
    color=orange,
    only marks, 
    mark=*  ,
     mark size=5pt
]  coordinates {
        (6, .700)
        (6, .650)
        (6, .872)
        (6, .821)
    };
    
    \node[anchor=north,color=red ,font=\footnotesize] at (axis cs:1,.85) {AI-TK1};
    \node[anchor=north,color=red, font=\footnotesize] at (axis cs:1, .745) {AI-TK2};
    \node[anchor=north,color=red, font=\footnotesize] at (axis cs:1, .55) {AI-TK3};
    \node[anchor=north,color=red, font=\footnotesize] at (axis cs:1, .46) {AI-TK4};
    \node[anchor=north, font=\footnotesize] at (axis cs:2, .83) {AI-TCK1};
    \node[anchor=north, font=\footnotesize] at (axis cs:2, .57) {AI-TCK2};
    \node[anchor=north, font=\footnotesize] at (axis cs:2, .646) {AI-TCK3};
    \node[anchor=south, font=\footnotesize] at (axis cs:2, .696) {AI-TCK4};
    \node[anchor=west,color=purple, font=\footnotesize] at (axis cs:3, .59) {AI-TPK1};
    \node[anchor=south,color=purple, font=\footnotesize] at (axis cs:3, .61) {AI-TPK3};
    \node[anchor=north,color=purple, font=\footnotesize] at (axis cs:3, .53) {AI-TPK4};
    \node[anchor=south,color=purple, font=\footnotesize] at (axis cs:3, .666) {AI-TPK5};
    \node[anchor=north,color=cyan, font=\footnotesize] at (axis cs:4, .67) {AI-TPACK1};
    \node[anchor=south,color=cyan, font=\footnotesize] at (axis cs:4, .6) {AI-TPACK2};
    \node[anchor=south,color=cyan, font=\footnotesize] at (axis cs:4, .7) {AI-TPACK3};
    \node[anchor=north,color=cyan, font=\footnotesize] at (axis cs:4, .56) {AI-TPACK5};
    \node[anchor=west, font=\footnotesize] at (axis cs:5, .514) {SE2};
    \node[anchor=north, font=\footnotesize] at (axis cs:5, .665) {SE3};
    \node[anchor=west, font=\footnotesize] at (axis cs:5, .806) {SE4};
    \node[anchor=west, font=\footnotesize] at (axis cs:5, .679) {SE5};
    \node[anchor=west,color=orange, font=\footnotesize] at (axis cs:6, .700) {TB1};
    \node[anchor=west,color=orange, font=\footnotesize] at (axis cs:6, .650) {TB2};
    \node[anchor=west,color=orange, font=\footnotesize] at (axis cs:6, .87) {TB4};
    \node[anchor=west,color=orange, font=\footnotesize] at (axis cs:6, .82) {TB5};
    \end{axis}
\end{tikzpicture}
\vspace*{-0.2cm}
\centering 
\caption{Exploratory factor analysis results (only factor loadings greater than 0.4 are shown).}
\footnotesize
\text{Note: SE represents Self-efficacy and TB represents Teaching Beliefs.} 
\label{factor}
\end{figure}
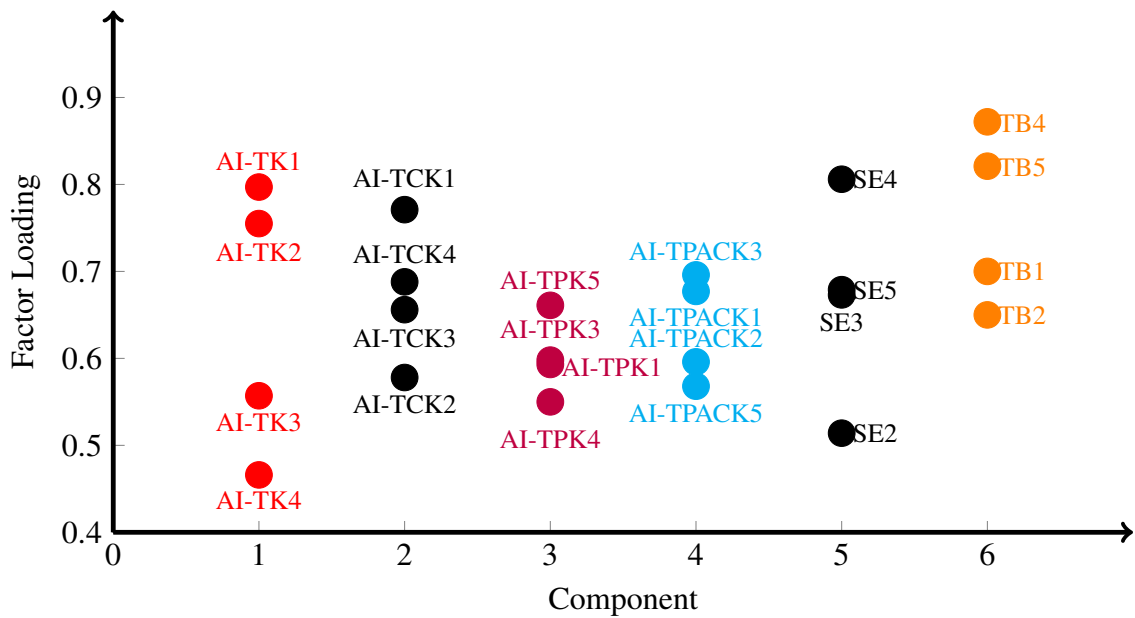

Before conducting EFA, it is necessary to ascertain the suitability of the data for factorisation. The Kaiser-Meyer-Olkin (KMO) was calculated to be 0.937, and the Barlett sphericity test yielded a significant result (3399.314, p \textless{} 0.001), indicating that the data set has the factor ability feature. Principal component analysis with maximum variance rotation was applied to explore the factor structure of the AI-TPACK scale. We found four items that were not part of the predetermined factors that needed to be removed, including AI-TK5, AI-TCK5, AI-TPK2, AI-TPACK4, SE1, TB3. The Fig.\ref{factor} indicated the factor loadings of the retained items. Factor loadings ranging from 0.466 to 0.879, indicating that all items and associated factors are robust. The result of the EFA was that the final version of the AI-TPACK scale contained 6 factors and 24 items. And Table \ref{tab:scale} showed descriptive statistics of all items. Furthermore, the total variance of the scale items is 83.87\%. The Cronbach’s alphas of all factors (AI-TK = 0.899, AI-TCK = 0.917, AI-TPK = 0.934, AI-TPACK = 0.930, Self-efficacy = 0.909, and Teaching Beliefs = 0.913) were discovered to be above the threshold value of 0.70, and, thus items were internally consistent within their factors.

\vspace*{-0.08cm}
\setlength{\extrarowheight}{2.8pt}
\begin{longtable}[t!]{p{2.5cm}@{\hspace{0.2cm}} p{10cm}@{\hspace{0.2cm}} p{1cm}@{\hspace{0cm}} p{1cm}@{\hspace{0cm}} p{1cm}} 
  \caption{Descriptive statistics (M, SD), and reliabilities ($\alpha$) of the revised AI-TPACK scale (24 items).} \label{tab:scale} \\
  \toprule
  \textbf{Item} & & \textbf{M} & \textbf{SD} & \textbf{$\alpha$} \\
  \midrule
  \endfirsthead 

  \multicolumn{5}{c}%
  {{ \tablename\ \thetable{} -- Continued from previous page}} \\
  \toprule
  \textbf{Item} & & \textbf{M} & \textbf{SD} & \textbf{$\alpha$} \\
  \midrule
  \endhead 

  \midrule
  \multicolumn{5}{r}{{Continued on next page}} \\
  \endfoot 

  \bottomrule
  \endlastfoot 

    \textbf{AI-TK} & & & & 0.899 \\
    AI-TK1 & I am familiar with the AI tools available currently for mathematical teaching & 3.68 & 1.48 & \\
    AI-TK2 & I understand the potential applications of AI in mathematical education, such as personalised learning & 4.03 & 1.35 & \\
    AI-TK3 & I am able to use at least one AI tool for assistance in mathematical teaching or learning & 4.3 & 1.19 & \\
    AI-TK4 & I can teach students how to use AI tools to assist their mathematical learning effectively & 4.12 & 1.29 & \\
    \textbf{AI-TCK} & & & & 0.917 \\
    AI-TCK1 & I understand the unique pedagogical advantages of AI tools combined with specific mathematical content & 4.33 & 1.23 & \\
    AI-TCK2 & I know how to select AI tools to fit specific mathematical content & 4.13 & 1.26 & \\
    AI-TCK3 & I am actively involved in continuing education and training opportunities related to AI& 3.97 & 1.45 & \\
    AI-TCK4 & I follow the latest AI developments in mathematics education and try to integrate them into teaching & 4.21 & 1.28 & \\
    \textbf{AI-TPK} & & & & 0.934 \\
    AI-TPK1 & I understand the capabilities and limitations of AI tools in dealing with specific mathematical problems, such as geometric proving, equation solving & 4.14 & 1.26 & \\
    AI-TPK3  & I can identify which mathematical concepts, propositions, problem solving or skills could be taught more effectively with AI tools & 4.17 & 1.20 & \\
    AI-TPK4  & I know how to use AI tools to create interactive and collaborative mathematical learning environments & 4.22 & 1.26 & \\
    AI-TPK5  & I can choose the right AI tools to help students understand abstract mathematical concepts and propositions & 4.26 & 1.24 & \\
    \textbf{AI-TPACK} &  & & & 0.93 \\
    AI-TPACK1 & I can combine AI tools with mathematics content and methods effectively to achieve my teaching objectives & 4.27 & 1.25 &\\
    AI-TPACK2 &  When designing a mathematical course, I can integrate AI tools to improve students' learning effectiveness & 4.26 & 1.17 &\\
    AI-TPACK3 & I use AI tools to assess students' learning progress and understanding of mathematics & 4.15 & 1.17 &\\
    AI-TPACK5 & I can effectively combine AI with a variety of assessment methods to evaluate students' learning outcomes & 4.20 & 1.30 &\\
    \textbf{SE} & & & & 0.909\\
    SE2 & I believe I can effectively use AI tools to improve my teaching efficiency & 4.46 & 1.24 &\\
    SE3 & I think I can use AI flexibly in math class to suit different students' learning styles & 4.23 & 1.19 &\\
    SE4 & I am confident in using AI to conduct mathematics research to enhance my professional knowledge and teaching skills & 4.42 & 1.21 &\\
    SE5 & I think I can effectively use AI tools to improve students' mathematical skills, such as problem-solving & 4.39 & 1.29 &\\
    \textbf{TB} & & & & 0.913 \\
    TB1 & I firmly believe that AI is an important tool for improving the effectiveness of mathematics teaching and will be integrated into teaching practice & 4.58 & 1.17 &\\
    TB2 & I attach great importance to the use of AI to support individualized instruction & 4.52 & 1.25 &\\
    TB4 & I believe that mathematics teaching should incorporate AI tools to enhance students' learning experience & 4.80 & 1.21 &\\
    TB5 & I believe that AI should be used in mathematics teaching to better reflect the close the link with real life & 4.82 & 1.23 &\\
\end{longtable}

We then conducted CFA using AMOS 24 to test the theoretical structure of AI-TPACK and to validate AI-TPACK scale. The CFA results and assessments of the model are shown in Table \ref{tab: criterion}, the result meet most metrics criterion \citep{Hu1999data}. And correlated residuals of all latent variables are above 0.68. It was shown that the fit of our proposed AI-TPACK-SEM is satisfactory, thus demonstrating the construct validity of the AI-TPACK measures.

\vspace*{-0.08cm}
\begin{table*}[!ht]
 \caption{The confirmatory factor analyze results and assessments of the model.}
 \label{tab: criterion}
    \centering
    \small
\scalebox{1.0}
    {    
    \setlength{\extrarowheight}{3pt}
    \begin{tabular}{p{2.5cm}@{\hspace{0cm}} >{\centering\arraybackslash}p{2.5cm}c >{\centering\arraybackslash}p{2.5cm}c >{\centering\arraybackslash}p{2.5cm}c >{\centering\arraybackslash}p{2.5cm}c}
    \toprule
    \textbf{Metrics}  & {Acceptable} & {Good} & {Results} & {Evaluation of results}
     \\
    \midrule
    $x^2$/$df$ & (3, 5] & $\leq$ 3 & 1.762 & Good\\
    RMSEA & (0.05, 0.08] & $\leq$ 0.05 & 0.077 & Acceptable\\
    NFI & (0.90, 0.95] & \textgreater 0.95 & 0.903 & Acceptable\\
    CFI & (0.95, 0.97] & \textgreater 0.97 & 0.955 & Acceptable\\
    TLI & (0.90, 0.95] & \textgreater 0.95 & 0.913 & Acceptable\\
    AGFI & (0.85, 0.90] & \textgreater 0.90 & 0.767 & Close to Acceptable \\
    GFI & (0.85, 0.90] & \textgreater 0.90 & 0.833 & Close to Acceptable\\
    \bottomrule
    \end{tabular}
}
    \vspace*{0.1cm}
    \vspace*{-0.1cm}
\end{table*}

\subsection{Post-test}
\label{4.2}
\subsubsection{Descriptive
statistical analyses}
\label{4.2.1}
\begin{figure}[ht!]
\centering
\includegraphics[width=0.6\textwidth]{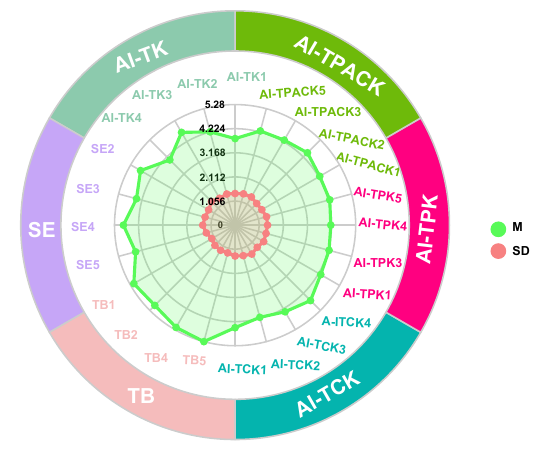}
\caption{Descriptive statistics (M, SD) of six variables.}
\label{mvalue}
\end{figure}

 Firstly, questionnaires were distributed via the internet to students at three grade levels from seven universities. We finally received 412 valid questionnaires. The mean value and standard deviation (SD) of the six variables are shown in Fig.\ref{mvalue}. The mean values are range from 3.7 to 5.3, which are all at the medium level. However, a comparison of the six variables reveals that teaching beliefs and self-efficacy exhibit higher levels than the AI-TPACK components (AI-TK, AI-TCK, AI-TPK, and AI-TPACK). The mean value of teaching beliefs (M=5.18) is the highest, while that for AI-TK (M=4.18) is the lowest. With regard to AI-TPACK components, the mean of AI-TCK (M=4.44) is the most elevated. However, in general, the discrepancy is minimal.

\begin{figure}[ht!]
\centering
   \includegraphics[width=0.9\textwidth]{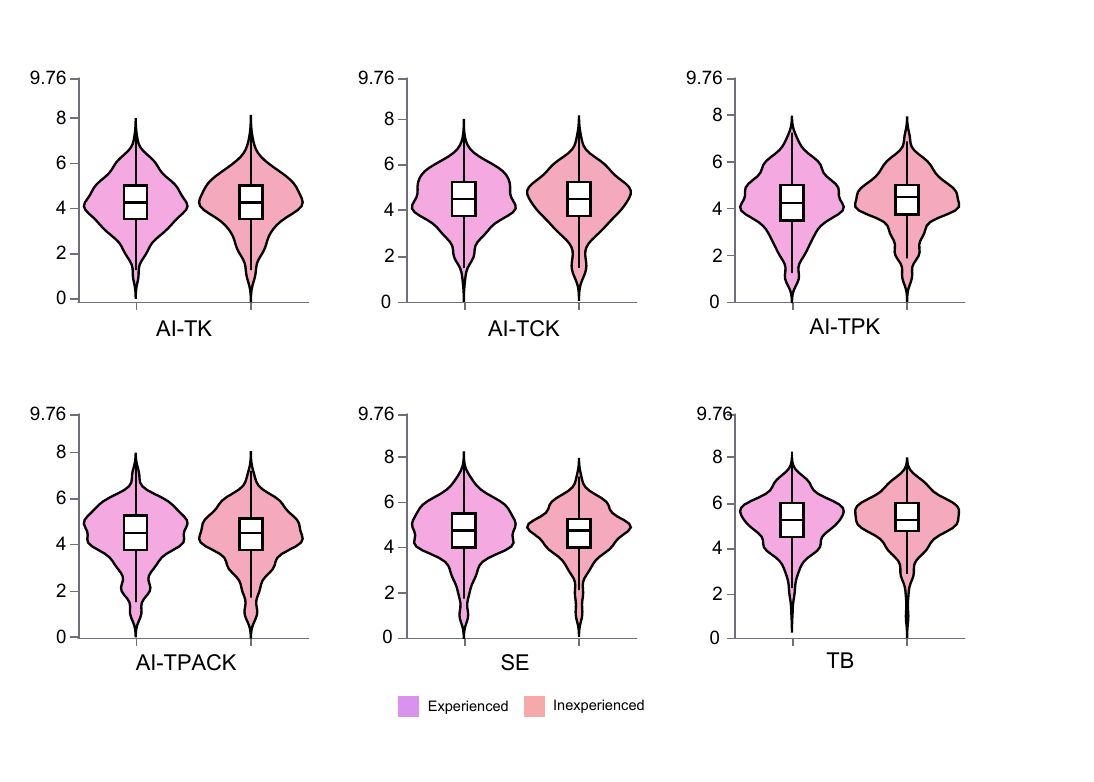}
\caption{Differences in scores on the six variables between experienced and inexperienced participants.}
\label{experienced}
\end{figure}

Then, a further test was conducted to ascertain whether there were differences between experienced MTES and those without experience in their responses to the AI-TPACK. Experienced in this context refers to a period of teaching practice in an educational institution. As illustrated in Fig.\ref{experienced}, the mean scores of experienced MTES were slightly higher than those inexperienced on every variables. It has been demonstrated that teaching experience does not significantly enhance teachers' proficiency in AI-TPACK.

\begin{figure}[ht!]
\centering
\includegraphics[width=0.6\textwidth]{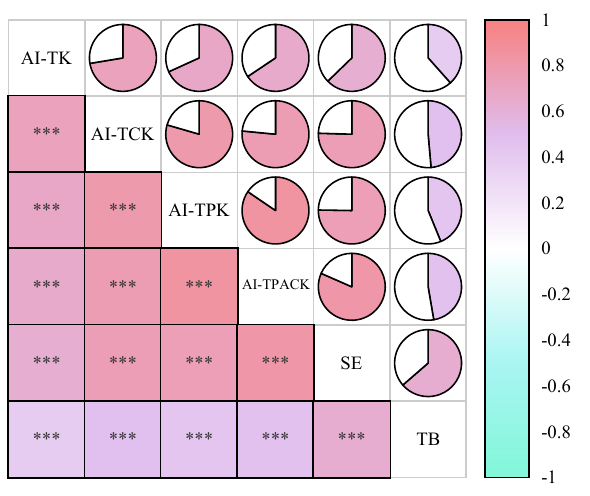}
\caption{Correlation matrix of six variables}
\label{correction}
\end{figure}

\begin{figure}[ht!]
\centering
\includegraphics[width=1\textwidth]{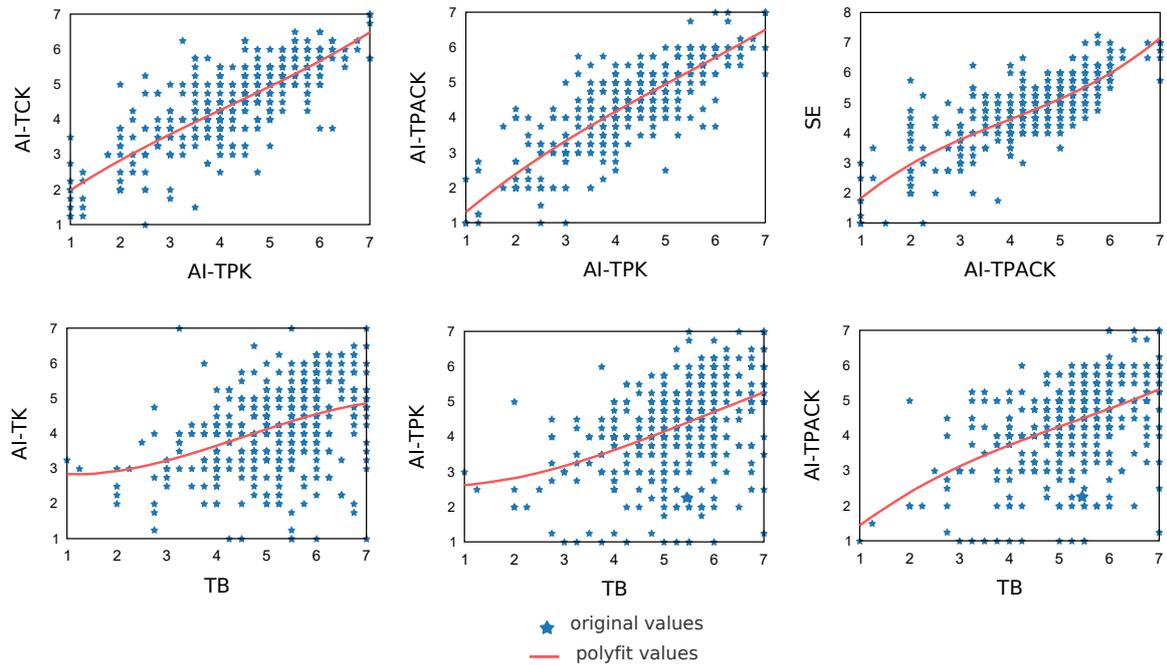}
\caption{Detailed correlation between two of the six variables}
\label{detail}
\end{figure}

 As illustrated in the Fig.\ref{correction}, the results of Pearson's correlation analysis indicate that all six variables are correlated to some extent. It can be observed that there is a robust correlation between the constituent elements of the AI-TPACK. However, the direct correlation between AI-TK and AI-TPACK is relatively weak, and the correlation with the AI-TPACK can be strengthened by AI-TCK and AI-TPK. Three strongly correlated and three weakly correlated relationships were selected for further analysis, with detailed distributions of scores and approximate curve fitting between two of the six variables are provided in Fig.\ref{detail}. In the scatter plot of AI-TPK versus AI-TCK and AI-TPACK, the blue dots are observed to be positioned in a relatively central location above and below the red fitted line, which suggests that there is indeed a strong linear relationship between the two variables. Especially the scatter plot of AI-TPACK and self-efficacy displays a strong fit, indicating a mutual reinforcement between the two. Self-efficacy exerts an extremely robust facilitating effect on AI-TPACK. However, TB keeps a relatively weak correlation with the internal elements AI-TPACK, particularly in the case of AI-TK, AI-TCK, and AI-TPACK. The blue dots in the figure exhibit greater dispersion, and the red curve displays a flatter profile, indicating a diminished impact.

\subsubsection{Differences within and between grades}
\label{4.2.2}
\begin{figure}[ht]
\centering
\includegraphics[width=1\textwidth]{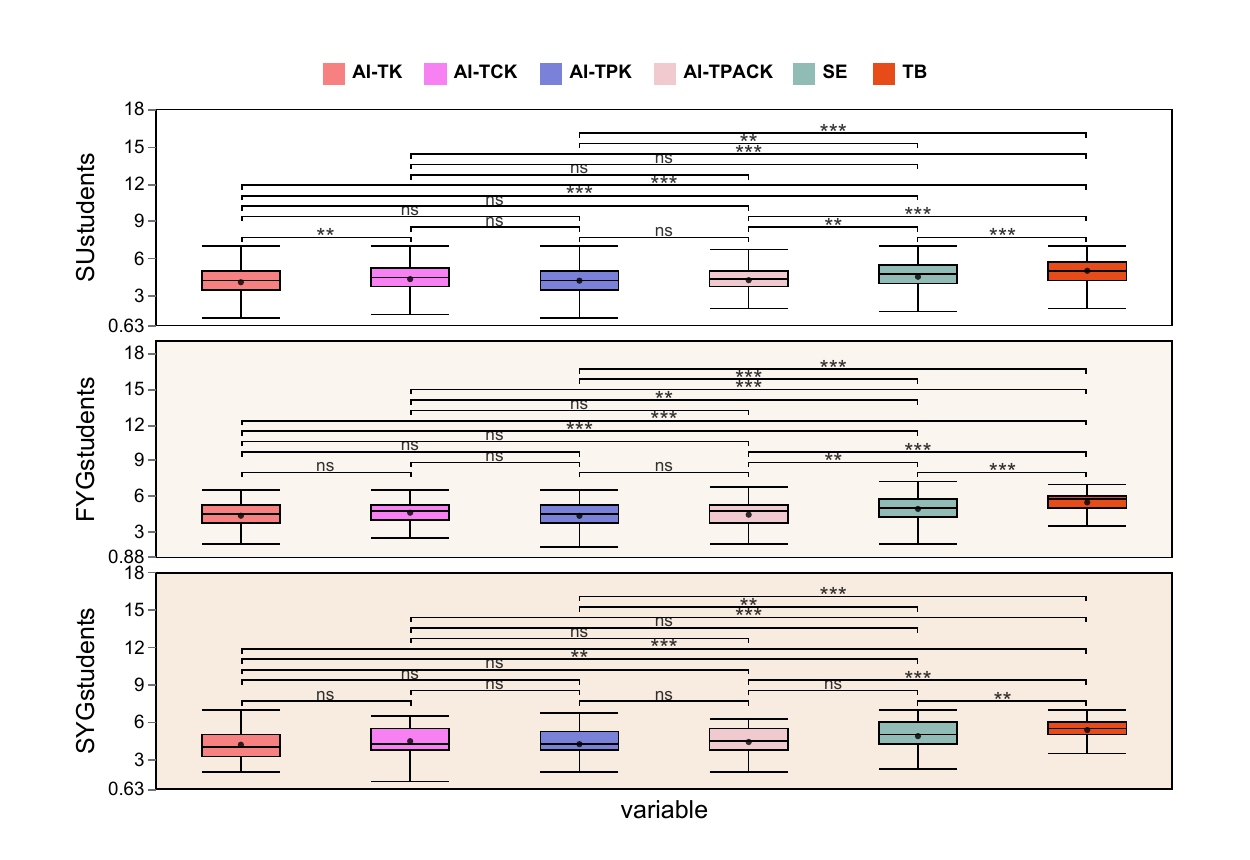}
\caption{Comparison of six variables within the three different grade levels.}
\footnotesize
\noindent\begin{minipage}{\textwidth}
    \textbf{Note:} SUstudents = senior university students, FYGstudents = first-year graduate students, 
    SYGstudents = second-year graduate students, ns = not significant, **p \textless{} .01, ***p \textless{} .001
    \end{minipage}
\label{within}
\end{figure}

In order to gain insight into the differences of variability, we initially conducted a comparison  between the six variables within each grade level. The results are presented in the Fig.\ref{within}. In senior university students, it indicated that significant differences between AI-TK and AI-TCK (p\textless{}0.01), self-efficacy (p\textless{}0.005), teaching beliefs (p\textless{}0.005). In first-year graduate students, while there are no significant differences between the elements within AI-TPACK, there are significant differences with both self-efficacy and teaching beliefs. First-year graduate students and second-year graduate students exhibited a high degree of similarity, with the exception of the observation that AI-TCK and AI-TPACK did not demonstrate a statistically significant difference from self-efficacy. It is noteworthy that significant differences in self-efficacy and teaching beliefs were observed in all three grades.

Then, in order to ascertain the extent of the differences in the six variables across the grades, an ANOVA process was conducted and Sig \textless{} 0.05 means there exist significance differences. As illustrated in Table \ref{tab: anova}, AI-TK (F=1.83, Sig=0.161), AI-TCK (F=1.90, Sig=0.151), AI-TPK (F=0.37, Sig=0.708) and AI-TPACK (F=0.83, Sig=0.435) have no significant difference between grades. Significance differences existed in Self-efficacy (F=4.68, Sig=0.010) and Teaching Beliefs (F=7.58, Sig=0.001).

\vspace*{-0.08cm}
\begin{table}[!ht]
 \caption{Differences between six variables in the three grade levels.}
 \label{tab: anova}
    \centering
    \small
    \footnotesize
\scalebox{1.0}
    {    
    \setlength{\extrarowheight}{5pt}
    \begin{tabular}{l | >{\centering\arraybackslash}p{1.2cm} >{\centering\arraybackslash}p{1.2cm} | >{\centering\arraybackslash}p{1.2cm} >{\centering\arraybackslash}p{1.2cm} | >{\centering\arraybackslash}p{1.2cm} >{\centering\arraybackslash}p{1.2cm} | >{\centering\arraybackslash}p{1.2cm} >{\centering\arraybackslash}p{1.2cm}}
    \toprule
    \multirow{2}{*}{\textbf{Variable}} & \multicolumn{2}{c}{\textbf{SUstudents}} & \multicolumn{2}{c}{\textbf{FYGstudents}} & \multicolumn{2}{c}{\textbf{SYGstudents}} & \multirow{2}{*}{F}  & \multirow{2}{*}{Sig}
    \\
    \cmidrule(lr){2-3} \cmidrule(lr){4-5} \cmidrule(lr){6-7} 
     & M & SD & M & SD & M & SD & & \\
    \midrule
    \text{\textbf{AI-TK}} &4.11&1.16&4.37&1.17&4.19&1.25&1.83&0.161 \\
    \text{\textbf{AI-TCK}} &4.36&1.19&4.61&1.03&4.48&1.22&1.90&0.151 \\
    \text{\textbf{AI-TPK}} &4.24&1.32&4.36&1.21&4.24&1.30&0.37&0.708 \\
    \text{\textbf{AI-TPACK}} & 4.27&1.29&4.45&1.19&4.41&1.29&0.83&0.435 \\
    \text{\textbf{Self-efficacy}} & 4.55&1.24&4.94&1.07&4.88&1.20&4.68&0.010 \\
    \text{\textbf{Teaching Beliefs}} & 5.03&1.12&5.49&0.94&5.39&1.05&7.58&0.001\\
    \bottomrule
    \end{tabular}
    }
    
    \vspace{0.5em}
    \centering 
    \footnotesize
    \begin{minipage}{\textwidth}
    \centering
    \textbf{Note:} M = mean value, SD = standard deviation, F = F-statistic, Sig = significance level.
    \end{minipage}
\end{table}

\subsubsection{AI-TPACK-SEM parsing}
\label{4.2.3}
We added two latent variables, self-efficacy and teaching beliefs, into the AI-TPACK and proposed a new AI-TPACK-SEM, as shown in the Fig.\ref{hypothesize-circle}. The aim is to elucidate the impact of the AI-TPACK components and these two latent variables on the development of AI-TPACK competence in MTES. And to understand whether self-efficacy and teaching beliefs could contribute to the development of AI-TPACK. 

\begin{figure}[ht!]
\centering
\includegraphics[width=0.8\textwidth]{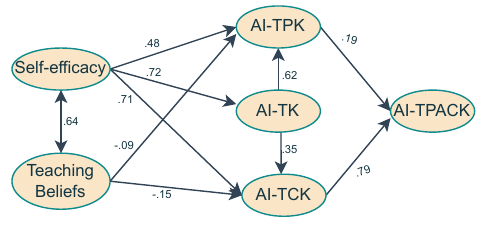}
\caption{The path parsing of AI-TPACK-SEM.}
\label{path}
\end{figure}

AI-TPACK-SEM analysis initially yielded both significant and insignificant relationships among the AI-TPACK components (AI-TK, AI-TCK, AI-TPK, and AI-TPACK) and Self-efficacy and Teaching Beliefs. We deleted insignificant relationships (H2b, H3c) and the model had an adequate fit to the data: ($x^2$/$df$) = 3.344, p \textless{} 0.001, RMSEA = 0.076, NFI = 0.908, CFI = 0.933, AGFI = 0.819, TLI = 0.923. As shown in Fig.\ref{path}, Self-efficacy was positively associated with AI-TPK ($\beta$ = 0.48), AI-TK ($\beta$ = 0.72) and AI-TCK ($\beta$ = 0.71). Hence, H1a, H1b and H1c were accepted. However, teaching beliefs was not positively with AI-TPK ($\beta$ = -0.09) and AI-TCK ($\beta$ = -0.15), H2a and H2c were rejected. AI-TK was positively associated with AI-TPK ($\beta$ = 0.62) and AI-TCK ($\beta$ = 0.35). Hence, H3a and H3b were accepted. We also revealed that both AI-TPK ($\beta$ = 0.19) and AI-TCK ($\beta$ = 0.79) were positively related to AI-TPACK, supporting H4 and H5.

\section{Discussion}
\label{sec5}
\subsection{The current status of AI-TPACK for MTES}
\label{5.1}
The results of the scale study indicated that all aspects of AI-TPACK for MTES were at the medium level. This suggests that they are generally willing to utilise AI tools or systems and hold a favourable attitude towards its application in education. However, the mean of AI-TK is the lowest, this indicate that MTES have not yet achieved proficiency in the utilisation of AI. As evidenced in Table \ref{tab: tool}, the AI tool most frequently utilized by students is the Question-and-Answer System (N=327) and Intelligent Tutoring Systems (N=218). They utilise the Question-and-Answer System to pose questions when they encounter a problem, with the objective of identifying a solution. For example, ChatGPT is employed to assist in formulating queries for examination papers, the creation of instructional designs, and other similarly creative tasks. In general, teaching support systems are designed in accordance with the specifications set forth by educational institutions. They facilitate the correction of homework tasks and enable educators to deliver personalised learning for students. Additionally, they assist in identifying areas of weakness or gaps in students' knowledge. Nevertheless, the utilisation of other tools (Intelligent Identification Observation System, Virtual Reality Technology and Educational Robotics) is minimal or absent in the context of other tools. However, these AI tools have the potential to enhance teacher trainee AI-TPACK and facilitate teaching in a more effective manner.

\vspace*{-0.08cm}
\begin{table*}[!ht]
 \caption{The utilisation of AI tools (\textbf{N} is the number of users).}
 \label{tab: tool}
    \centering
    \small
\scalebox{1.0}
    {    
    \setlength{\extrarowheight}{3pt}
    \begin{tabular}{p{7.5cm}@{\hspace{0cm}} p{7.5cm}@{\hspace{0cm}} >{\centering\arraybackslash}p{0.5cm}c}
    \toprule
    \textbf{Type of AI Tools} & {\textbf{Examples}}
     & \textbf{N}
     \\
    \midrule
    Intelligent Teaching System & FiF (\textit{\url{https://www.fifedu.com/}})& 166 \\
    Question-and-Answer System & ChatGPT (\textit{\url{https://chatgpt.com/}}) & 327 \\
    Intelligent Tutoring Systems & Zhixue (\textit{\url{https://www.zhixue.com/}})  & 218 \\
    Intelligent Linguistic Text Recognition System  & WPS AI (\textit{\url{https://www.wps.ai/}}) & 157 \\
    Intelligent Identification Observation System & EYESO (\textit{\url{http://www.eyeso.net/}}) & 6 \\
    Virtual Reality Environment & Smart Class & 55 \\
    Virtual Reality Technology & VR glasses & 36 \\
    Educational Robotics & Alpha egg (\textit{\url{http://www.toycloud.com/}})& 8 \\
    \bottomrule
    \end{tabular}
}
    \vspace*{0.1cm}
    \vspace*{-0.1cm}
\end{table*}

While the overall AI-TPACK level of MTES is at the medium level, they predominantly utilise only elementary and fundamental tools and have not yet attained the advanced AI proficiency demanded by the AI-TK. This makes it challenging for students to fully engage with the integration of AI in education, hindering the creation of an immersive teaching and learning experience. So the current status of AI-TPACK for MTES in China is at a basic, preliminary stage.

\textbf{Advices:} In order to address the issue at hand, it is imperative that MTES not only utilise the capabilities of ChatGPT in order to identify a solution, but also apply artificial intelligence technology to the field of teaching \citep{kim2022learning}. For example, the integration of virtual reality tools, which can facilitate a deeper understanding of the changes in geometric shapes. Additionally, the implementation of smart classrooms can provide a simulated environment that reflects the reality of utilising mathematical principles to address practical problems. It is similarly crucial for educational establishments to furnish their students with the requisite devices and environments, thereby facilitating the optimisation of AI-TPACK capabilities amongst those MTES to become outstanding teachers.

\subsection{Differences in MTES between grades}
\label{5.2}
The ANOVA indicates that there is no significant AI-TPACK difference between grades of MTES. One might reasonably inquire as to why there are no difference exists in AI-TPACK, but has difference in self-efficacy and teaching beliefs. The findings indicate that graduate education only led to an increase in their self-efficacy and teaching beliefs, and did not improve their AI-TPACK abilities. The field of mathematics is inherently abstract, and artificial intelligence (AI) can assist in visualizing these complex problems, thereby making them more approachable and facilitating deeper comprehension. So there is a distinction in the self-efficacy and teaching beliefs of MTES at different grades when using AI tools for the teaching of mathematics.

In conjunction with the findings presented in Table \ref{tab: way}, it is determined that the primary source of information for MTES regarding AI tools is the Internet (N=394). The acquisition of knowledge and proficiency in the application of AI through independent internet-based learning is not a systematic process. Furthermore, this approach is ineffective in enhancing the AI-TK, AI-TPK and AI-TCK of MTES, which is a significant obstacle to overall improvement. It is therefore imperative that educational establishments implement a structured curriculum similar like that incorporates artificial intelligence and pedagogical practices. Insufficient information is available on family and friends (N=139) involved. There is also a lack of communication and sharing of information about AI. It is noteworthy that the majority of respondents indicated that their internships (N=122) had provided the least valuable insight into the use of AI tools. This indicates that the teaching of secondary school students in China is still based on traditional methods and rarely incorporates the use of AI tools. This is a significant impediment to the digital development of education.

\vspace*{-0.08cm}
\begin{table*}[!ht]
 \caption{The way to know about AI tools (\textbf{N} is the number of participates).}
 \label{tab: way}
    \centering
    \small
\scalebox{1.0}
    {    
    \setlength{\extrarowheight}{1pt}
    \begin{tabular}{c c c c c c }
    \toprule
    \textbf{Ways}
     & {Internet} & {Family and Friends}& {Curriculum Studies}& {Specialised Lectures}& {Internships}
     \\
    \midrule
    \textbf{N} & 394&139&207&134&122 \\
    \bottomrule
    \end{tabular}
}
    \vspace*{0.1cm}
    \vspace*{-0.1cm}
\end{table*}

\textbf{Advices:} In light of the aforementioned analysis, it becomes evident that a restructuring of the university curriculum is imperative. The curriculum is structured in accordance with the characteristics of mathematics and teaching objectives, thereby facilitating the achievement of visualisation. On the one hand, it is recommended that the introduction and training of AI technology courses be strengthened in order to provide MTES with a more specialised understanding of AI-TK \citep{SOUTHWORTH2023100127}. As an illustration, this may include guidance on how to utilise AIGC in a more effective manner to obtain information. Furthermore, it would be advantageous to them on how to integrate AI-TK with CK, PK , to facilitate the advancement of AI-TCK and AI-TPK, thereby advancing the holistic development of AI-TPACK. On the other hand, the school curriculum should be differentiated according to grade level, with the objective of enabling students in higher grades to develop greater proficiency in AI-TPACK \citep{HORNBERGER2023curriculum}. For instance, the curriculum for first-year graduate students places greater emphasis on the utilisation of AI tools, whereas second-year graduate students are required to demonstrate a deeper understanding of how these tools can be effectively integrated into the teaching process. And the curriculum should facilitate the establishment of a learning community and the implementation of peer-assisted learning driven by project tasks, thereby enhancing the development of MTES' AI-TPACK competence.

\subsection{The effects of self-efficacy and teaching beliefs on AI-TPACK}
\label{5.3}
In the context of teaching, self-efficacy and teaching beliefs are of significant importance. By SEM, it is possible to analyse how self-efficacy and teaching beliefs affect AI-TK, AI-TCK, AI-TPK. A positive correlation was observed between self-efficacy and AI-TK (H1b), AI-TCK (H1c) and AI-TPK (H1a). This indicates that an increase in self-efficacy is conducive to the comprehensive advancement of AI-TPACK. It can be posited that an increase in self-efficacy among MTES will result in a corresponding increase in confidence when applying AI-TK in their teaching practice. This, in turn, will enhance the effectiveness of their teaching. Furthermore, they demonstrate enhanced confidence in their capacity to address the challenges they encounter in the utilisation of AI tools.

Unexpectedly, teaching beliefs affected AI-TCK (H2c) and AI-TPK (H2a) negatively and had no significant effect on AI-TK (H2b). This suggests that an increase in MTES' teaching beliefs will result in greater conviction in the effectiveness of traditional teaching methods. And the mean value of teaching beliefs (M=5.18) was the highest, while that for AI-TK (M=4.18) was the lowest. Consequently, there is a possibility that the utilisation of AI technology will encounter resistance, which may impede the advancement of AI-TCK and AI-TPK competencies. The pedagogical characteristics of mathematics education are characterised by a paucity of instructional time in comparison to the volume of content that needs to be covered, as well as a pragmatic focus on examination outcomes. Furthermore, the prevalence of traditional teaching methodologies, which are sufficient for routine instructional purposes, has resulted in a notable reluctance among MTES to explore the integration of more AI tools in their teaching practices. Despite a prevailing positive attitude towards AI, there is a notable dearth of enthusiasm for its utilisation in the field of education. 

Contrary to our hypothesis, AI-TK was associated with AI-TPK (H3a) and AI-TCK (H3b), but not with AI-TPACK (H3c). This indicates that AI-TK is unable to contribute directly to the advancement of AI-TPACK. Instead, it must be incorporated into PK, CK as a means of fostering the growth of AI-TPK and AI-TCK. And subsequently, AI-TPK and AI-TCK  facilitate AI-TPACK (H4, H5). This demonstrates that the elements of AI-TPACK are interrelated and mutually reinforcing, and that the advancement of any given aspect can contribute to the enhancement of the whole.

\textbf{Advices:} The evidence presented here demonstrates that teachers' self-efficacy plays a significant role in the advancement of AI-TPACK. Consequently, it is crucial to cultivate the self-efficacy of MTES. For instance, it would be beneficial to assist them in enhancing their self-confidence and willingness to utilise AI tools in the event of encountering difficulties \citep{nazaretsky2022trust}. And facilitating the integration of AI technology into their pedagogical practice to attain optimal outcomes fosters the growth of both self-efficacy and AI-TPACK. Furthermore, the weaker negative correlation between teaching beliefs and AI-TPACK necessitates a balanced approach to their relationship. In response, it is crucial to assist MTES in updating their teaching beliefs, thereby enhancing their responsiveness to the pedagogical requirements of AI-TPACK. It is incumbent upon school leadership to develop policies that support technology integration, provide the necessary resources and training, and foster a school culture that encourages innovation and experimentation with new technologies. At the same time, a uniform set of standards for the evaluation of teaching internships for MTES should be established, with the objective of facilitating their ability to effectively engage with the pedagogical implications of AI, as well as to keep pace with the evolving landscape of educational technology \citep{choi2023beliefs}. In this manner, not only are their self-efficacy and teaching beliefs enhanced, but also are their competencies in AI-TPACK optimized. It is only then that the digital transformation of education can be more favourably promoted.

\section{Conclusion}
\label{sec6}
In conclusion, in order to gain insight into the current state of AI-TPACK competence among MTES in China. We designed a technical framework for studying the AI-TPACK of MTES and devised an AI-TPACK scale for MTES. Through the pre-test, we ascertained the scale's reliability and validity. Then the implementation of the post-test yielded the following findings.

Firstly, our research revealed that the participants demonstrated at the medium level, yet they were utilising AI tools that were essentially rudimentary, indicating that they were still in the nascent stage of development. Secondly, we compared MTES from three different grades on the six variables (AI-TK, AI-TCK, AI-TPK, AI-TPACK, Self-efficacy, Teaching Beliefs), but found that there is no discernible difference between MTES from different grades, which suggested that a restructuring of
the university curriculum is imperative. Thirdly, a new AI-TPACK-SEM was designed to explain the relationship between AI-TPACK and self-efficacy and teaching beliefs. We come to a conclusion that may be contrary to common perception, excessive teaching beliefs may impede the advancement of AI-TPACK. This finding says it is important to ensure a harmonious and mutually reinforcing relationship between the two domains, that of teaching and learning, so that neither becomes unduly dominant and impedes the development of AI-TPACK. We hope that the research framework, findings and recommendations presented in this paper will prove invaluable to subsequent research and the integration of AI and TPACK, thereby facilitating the advancement of digital education.

It should be noted that the study is subject to certain limitations. Firstly, the data for this study were collected using validated scale, which, while ensuring both validity and reliability, still require the use of multiple data modalities. It would be beneficial for future research to combine qualitative data obtained through interviews or videos with quantitative data in order to gain a more comprehensive understanding of MTES' AI-TPACK from multiple perspectives.
Secondly, the relatively small number of second-year graduate students participating in the survey resulted in an under-representation of this group. It is therefore recommended that future studies should aim to include a larger number of students in order to enhance the reliability of the findings.

\section*{Data availability}
The research data associated with this paper can be accessed at \url{https://drive.google.com/drive/folders/1jGfAJpv6nptkJouZoMU6CY6x4IRvnKVt?usp=sharing}.

\section*{Conflict of interest disclosure} 
The authors declare no competing interests.




\end{document}